\documentclass[aps,prd,reprint]{revtex4-1}
\usepackage{epsfig}
\usepackage{amssymb}
\usepackage{amsthm}
\usepackage{amsmath}
\usepackage{graphicx}
\usepackage{color}
\usepackage{slashed}
\usepackage{mathrsfs}
\usepackage{float}
\usepackage{color}

\newcommand{\p}{\partial}

\newcommand{\MeV}{\;\text{MeV}}

\definecolor{red}{rgb}{1,0,0}

\begin{document}

\title{Chiral phase transition and meson spectrum in improved soft-wall AdS/QCD}

\author{Zhen Fang}
\email{fangzhen@itp.ac.cn}
\affiliation{Key Laboratory of Theoretical Physics, Institute of Theoretical Physics, \\
Chinese Academy of Science, Beijing 100190, P. R. China}
\affiliation{University of Chinese Academy of Sciences, Beijing, P.R. China}
\author{Yue-Liang Wu}
\email{ylw@itp.ac.cn}
\affiliation{Key Laboratory of Theoretical Physics, Institute of Theoretical Physics, \\
Chinese Academy of Science, Beijing 100190, P. R. China}
\affiliation{University of Chinese Academy of Sciences, Beijing, P.R. China}
\author{Lin Zhang}
\email{zhanglin@itp.ac.cn}
\affiliation{Key Laboratory of Theoretical Physics, Institute of Theoretical Physics, \\
Chinese Academy of Science, Beijing 100190, P. R. China}
\affiliation{University of Chinese Academy of Sciences, Beijing, P.R. China}




\begin{abstract}
We investigate in detail the chiral thermal transition of QCD in an improved soft-wall AdS/QCD model with a simply modified 5D conformal mass of the bulk scalar field. We also present a calculation in this model for the light meson spectra and other low-energy characteristic quantities including the pion form factor, the $\pi-\rho$ coupling constant and the decay constants of $\pi$, $\rho$, $a_1$, which are shown to result in a good agreement with experimental data except for the pion decay constant. The thermal behavior of chiral condensate is studied. It is found that such a simply improved soft-wall model incorporates the crossover behavior of chiral thermal transition indicated by lattice simulations. The expected chiral transition temperature can be obtained.
\end{abstract}
\maketitle


\section{Introduction}
\label{Introduction}
Quantum chromodynamics (QCD) in the low energy region is still not well understood for its strong coupling nature, which entails nonperturbative approaches such as lattice QCD approach. Various of effective field theories have also been built to tackle the low energy problems of QCD, which is usually based on the approximate global chiral symmetry and dynamical chiral symmetry breaking \cite{Nambu:1960xd}. It has been shown that the chiral dynamic model with dynamically generated spontaneous chiral symmetry breaking only enables us to predict the mass spectra of all light ground state mesons \cite{DW}.

In recent years, the Anti-de Sitter/Conformal field theory (AdS/CFT) correspondence \cite{Maldacena:1997re,Gubser:1998bc,Witten:1998qj}, which establishes the duality between the weakly coupled supergravity in $\rm{AdS}_5$ and the $\mathcal{N}=4$ super Yang-Mills gauge theory in the boundary, has been developed and applied successfully in many fields relevant to strong couplings. Particularly, many models based on AdS/CFT have been constructed to handle the low energy nonperturbative problems of QCD, which is usually called holographic QCD or AdS/QCD. There are top-down and bottom-up approaches in this direction. In the top-down approach one utilizes certain brane construction in string theory to describe the low energy phenomena of QCD \cite{Kruczenski:2003uq,Sakai:2004cn,Sakai:2005yt}, while in the bottom-up approach the bulk theory is usually constructed and constrained on the basis of the fundamental features of low energy QCD such as the chiral symmetry breaking and confinement property.

In the bottom-up approach, many models have been built to characterize the low energy hadron physics, such as the hard-wall and soft-wall models \cite{DaRold:2005zs,Erlich:2005qh,Karch:2006pv} and the light-front AdS/QCD model \cite{BT1,Brodsky:2014yha}. The hard-wall model \cite{DaRold:2005zs,Erlich:2005qh} introduces a sharp cut-off in the extra dimension to mimic the confinement of QCD. The chiral symmetry breaking pattern can be well reproduced in this model; however, it cannot obtain the linear Regge behavior of hadron spectrum which is identified as a typical feature of QCD confinement. The original soft-wall model \cite{Karch:2006pv} was constructed to reproduce the correct Regge behavior of meson spectrum by introducing an infrared (IR) suppressed dilaton term, yet the chiral symmetry breaking phenomena cannot be realized consistently.

To realize the confinement of QCD and the chiral symmetry breaking consistently, nonlinear interactions of bulk scalar field have been considered in \cite{Babington:2003vm,Casero:2007ae,Iatrakis:2010jb,Jarvinen:2011qe,Gherghetta:2009ac,Sui:2009xe,Li:2012ay,Cui:2016ocl}, and a quartic term of bulk scalar has proved to be crucial both for a consistent description of meson spectrum \cite{Gherghetta:2009ac,Sui:2009xe,Cui:2016ocl} and for the correct behavior of chiral phase transition \cite{Chelabi:2015cwn,Chelabi:2015gpc}. There are also many works which considered modified metric forms or dilaton profiles etc \cite{Gherghetta:2009ac,Sui:2009xe,Li:2012ay,Cui:2016ocl}. In \cite{Cui:2016ocl}, we constructed an IR-improved soft-wall AdS/QCD model and calculated the light meson spectra which fit the experimental data quite well. In \cite{Chelabi:2015cwn,Chelabi:2015gpc}, we studied the temperature dependent behavior of the chiral condensate in a modified soft-wall model and obtained the correct behavior of chiral phase transition.

To incorporate the meson spectrum, in this paper we propose a simply improved AdS/QCD model based on the previous work \cite{Cui:2016ocl} and investigate the chiral thermal transition behavior following the studies in \cite{Chelabi:2015cwn,Chelabi:2015gpc}. The vacuum expectation value (VEV) of the bulk scalar will be solved directly from the equation of motion (EOM), so the chiral condensate will be given as a derived quantity, and the dynamical chiral symmetry breaking pattern can be realized naturally in the soft-wall framework. To give a consistent description of the low-energy QCD phenomenon, we need to introduce a modified 5D conformal mass of the bulk scalar field which is constrained by the well-motivated UV and IR asymptotics, as did in our previous work \cite{Cui:2016ocl}. Physically, it is reasonable to modify the 5D conformal mass of the bulk scalar field in consideration of the quark mass anomalous dimension and the mass-dimension relation of AdS/CFT. It will be seen that the correct behavior of chiral thermal transition indicated from lattice simulations can be obtained in our simply improved soft-wall AdS/QCD model when combined with the calculation of light meson spectra.

The paper will be organized as follows. In Sec.\ref{model}, we outline the improved soft-wall AdS/QCD model with a modified 5D mass of the bulk scalar field. In Sec.\ref{zero temperature}, we compute the light meson spectra and other low-energy characteristic quantities including the pion form factor, the $\pi-\rho$ coupling constant and the decay constants of $\pi$, $\rho$ and $a_1$ mesons. In Sec.\ref{chiraltransition}, the behavior of chiral phase transition is investigated in this simply improved soft-wall model. In Sec.\ref{conclusion}, we give a summary and conclusion of our work.

\section{The improved soft-wall AdS/QCD model for mesons}\label{model}

The modified soft-wall AdS/QCD model is defined in the $\mathrm{AdS}_5$ space with the metric ansatz:
\begin{equation}\label{metric}
ds^2=e^{2A(z)}\left(\eta_{\mu\nu}dx^{\mu}dx^{\nu}-dz^2\right),
\end{equation}
where $\eta^{\mu\nu}=\mathrm{diag}\{+1,-1,-1,-1\}$ and $A(z)=-\mathrm{log}\frac{z}{L}$ with $L$ the $\rm{AdS}$ curvature radius which will be set to one for simplicity in the calculation below.

Now we present the modified soft-wall AdS/QCD model. The meson sector of the 5D action can be written as
\begin{equation}\label{meson action}
\begin{split}
S_M=\frac{1}{k}\int d^{5}x\,\sqrt{g}\,e^{-\Phi(z)}\,{\mathrm{Tr}}\{&|DX|^{2}-m_5^2(z)|X|^{2}\\
&-\lambda |X|^{4}-\frac{1}{4g_{5}^2}(F_{L}^2+F_{R}^2)\},
\end{split}
\end{equation}
where $D^MX=\p^MX-i A_L^MX+i X A_R^M$, $F_{L,R}^{MN}=\partial^MA_{L,R}^N-\partial^NA_{L,R}^M-i[A_{L,R}^M,A_{L,R}^N]$,
$A_L^M=A_L^{a,M}t_L^a$, $A_R^M=A_R^{a,M}t_R^a$, $t_L^a$ and $t_R^a $ are the generators of $\mathrm{SU}(2)_L$ and $\mathrm{SU}(2)_R$ respectively, and $\Phi(z)=\mu_g^2\,z^2$ is the dilaton profile which is necessary for the Regge behavior of meson spectrum \cite{Karch:2006pv}. $g_5$ can be determined by comparing the large momentum expansion of the correlator of vector current $J_{\mu}^a=\bar{q}\gamma_{\mu}t^aq$ in both AdS/QCD and perturbative QCD \cite{Erlich:2005qh}. It should be noted that a factor $\frac{1}{k}$ outsides the integral has been set, following \cite{Colangelo:2008us}, which is contrast to the previous works \cite{Erlich:2005qh,Karch:2006pv,Gherghetta:2009ac,Sui:2009xe,Li:2012ay,Cui:2016ocl} where $k=1$ was usually taken. Later we will give some details of discussion for the values of $k$ and $g_5$.

The bulk scalar field $X$ and the chiral gauge fields $A_{L,R}^M$ are dual to relevant QCD operators at the boundary $z=0$ by the AdS/CFT dictionary \cite{Erlich:2005qh}. In general, the bulk scalar field $X$ can be decomposed into the pseudoscalar meson field $\pi(x,z)=\pi^a(x,z) t^a$ and the scalar meson field $S(x,z)$ in the form of $X=(\frac{\chi}{2}+S)e^{2i\pi}$, where $\chi(z)$ is related to the VEV of $X$ by $\langle X\rangle=\frac{\chi}{2}I_2$ with $I_2$ the $2\times2$ identity matrix. The bulk gauge fields can be recombined into the vector field $V^M=\frac{1}{2}(A_L^M+A_R^M)$ and the axial-vector field $A^M=\frac{1}{2}(A_L^M-A_R^M)$ with the transformed field strengths as follows
\begin{equation}
\small
\begin{split}
F_{V}^{MN}&=\frac{1}{2}(F_L^{MN}+F_R^{MN})   \\
&=\partial^MV^N-\partial^NV^M-i[V^M,V^N]-i[A^M,A^N],\\
\end{split}
\end{equation}
\begin{equation}
\small
\begin{split}
F_{A}^{MN}&=\frac{1}{2}(F_L^{MN}-F_R^{MN})    \\
&=\partial^MA^N-\partial^NA^M-i[V^M,A^N]-i[A^M,V^N].
\end{split}
\end{equation}

To fix the action of the improved soft-wall AdS/QCD model, we need to specify the form of the 5D mass $m_5^2(z)$ which is critical to a consistent description of both meson spectrum and chiral phase transition, as mentioned above. Theoretically, a $z$-dependent bulk scalar mass might originate from the quark mass anomalous dimension which can be related to $m_5^2$ by the AdS/CFT dictionary $m_5^2=(\Delta-p)(\Delta+p-4)$ with $\Delta$ the dimension of the $p$-form operator. We first investigate the UV and IR asymptotics of $m_5^2(z)$. The UV asymptotic expression of $m_5^2(z)$  can be obtained from the EOM of bulk scalar VEV which can be derived from the action (\ref{meson action}) as
\begin{equation}\label{eom-chi}
\begin{split}
&\chi''(z)+\left(3A'(z)-\Phi'(z)\right)\chi'(z)\\
&-e^{2A(z)}\left(m_5^2\,\chi(z)+\frac{\lambda}{2}\,\chi^3(z)\right)=0.
\end{split}
\end{equation}
According to the AdS/CFT dictionary \cite{Erlich:2005qh}, the bulk scalar VEV has the following behavior in the UV region:
\begin{eqnarray}\label{chi-UV}
\chi(z\sim 0)=m_q\,\zeta\,z+\frac{\sigma}{\zeta}\,z^3+\cdots
\end{eqnarray}
with $m_q$ the current quark mass, $\sigma$ the chiral condensate and $\zeta$ a normalization constant which will be interpreted below. Substituting the above asymptotic expression of $\chi(z)$ in Eq.(\ref{eom-chi}), the UV asymptotics of bulk scalar mass can be derived as
\begin{eqnarray}\label{m5-UV}
m_5^2(z\sim 0)=-3-(2\,\mu_g^2+\lambda\,m_q^2\,\zeta^2/2)\,z^2+\cdots,
\end{eqnarray}
where the leading constant term $-3$ can be determined from the AdS/CFT dictionary $m_5^2=(\Delta-p)(\Delta+p-4)$ by taking $p=0$ and $\Delta=3$, which is the dimension of the dual operator $\bar{q}_R q_L$ \cite{Erlich:2005qh}. Phenomenally, the IR asymptotics of $m_5^2(z)$ is constrained by the mass split of chiral partners, i.e., the vector and axial-vector mesons in the high excited states, which requires $\chi(z)$ to be linear in the IR region: $\chi(z\rightarrow\infty)=\mu_{\mathrm{IR}}\,z$. Inserting this relation in Eq.(\ref{eom-chi}), we get the IR expression of $m_5^2(z)$ as
\begin{eqnarray}\label{m5-IR}
m_5^2(z\rightarrow\infty) = -(2\,\mu_g^2+\lambda\,\mu_{\mathrm{IR}}^2/2)\,z^2-3+\cdots.
\end{eqnarray}
We find that the leading UV and IR expressions of $m_5^2(z)$ has the same form, both with a constant and a quadratic term. For simplicity, the following parameterization of $m_5^2(z)$ will be used:
\begin{eqnarray}\label{m5}
m_5^2(z)=-3-\mu_c^2\,z^2.
\end{eqnarray}

At last, we address the issue about the parameter-fixing of $k$ and $g_5$ in the action (\ref{meson action}). From the calculation of the correlator of vector current $J_{\mu}^a=\bar{q}\gamma_{\mu}t^aq$, one can get $k g_5^2=\frac{12\pi^2}{N_c}$ with $N_c=3$ \cite{Erlich:2005qh}. It was argued in \cite{Colangelo:2008us,Krikun:2008tf} that $k$ can be fixed by comparing the two-point correlation functions of scalar or pseudoscalar currents computed from AdS/QCD and the ones in perturbative QCD. However, the values of $k$ determined by these two ways are inconsistent with each other, e.g., k is determined as $k=\frac{16\pi^2}{N_c}$ by comparing the correlators of scalar current $\mathcal{O}_S^a=\bar{q}t^aq$ \cite{Colangelo:2008us}, while $k=\frac{9\pi^2}{2 N_c}$ by comparing the correlators of pseudoscalar current $\mathcal{O}_{\pi}=\bar{q}\gamma_5q$ \cite{Krikun:2008tf}. Putting aside the tension between the two ways of parameter-fixing which might need further consideration, in many previous studies we usually took $k=1$ and a normalization constant $\zeta=\frac{\sqrt{N_c}}{2\pi}$ in Eq.(\ref{chi-UV}) must be introduced to produce the correct $N_c$ scaling behavior of $m_q$ and $\sigma$ in this case \cite{Cherman:2008eh}, while in the cases with $k\neq 1$ we just set $\zeta=1$. In this work, we will also consider the case with $k=\frac{16\pi^2}{N_c}$ (we will call it case B) which was addressed in \cite{Colangelo:2008us} besides the usual case with $k=1$ (we will call it case A) and compare the results between these two cases, which might give us some guidance for further improvement in the framework of AdS/QCD.

\section{Numerical calculations at zero temperature}\label{zero temperature}
\subsection{Input parameters and bulk scalar VEV}\label{parameter-fit}
Before doing the calculation, we need to specify the four parameters $m_q$, $\mu_g$, $\mu_c$ and $\lambda$ by fitting meson spectrum or other measured quantities, and a physical analysis of these parameters might be helpful to the exposition of following studies. In the four parameters, the current quark mass $m_q$ should be tightly linked with the ground state pion mass due to the effect of explicit chiral symmetry breaking, and this is indeed the case. As $\mu_g$ is contained in the dilaton term which is crucial to the linear Regge behavior of meson spectrum, the value of $\mu_g$ is expected to be closely related to the $\Lambda_{\mathrm{QCD}}$ energy scale, i.e., $\mu_g \sim \Lambda_{\mathrm{QCD}}$. The parameter $\mu_c$ is introduced by the modified bulk scalar mass $m_5^2(z)$ which is necessary for a reasonable realization of chiral symmetry breaking and chiral phase transition, as will be shown below, so we might expect that the value of $\mu_c$ should be close to the energy scale of chiral symmetry breaking: $\mu_c\sim \Lambda_{\chi} \sim 1~\mathrm{GeV}$.

In our calculation, the parameter $m_q$ is fixed by the pion mass $m_{\pi}\simeq139\MeV$. As we will see below, the EOM of vector meson in our model includes only the parameter $\mu_g$, which could be fixed by fitting vector meson spectrum. The parameters $\mu_c$ and $\lambda$ determine the mass split of vector and axial-vector mesons indirectly by the bulk scalar VEV $\chi(z)$, and these two parameters could be fixed by a global fitting of the mass spectra of scalar and axial-vector mesons. The best fit of the four parameters in case A and case B is shown in Table \ref{parameter}.
\begin{table}
\begin{center}
\begin{tabular}{ccccc}
\hline\hline
Case&  $m_q$(MeV) & $\mu_g$(MeV) & $\mu_c$(MeV)  & $\lambda$ \\
\hline
   $A$ &    3.366    &   440        &     1180      &    33.6  \\
\hline
   $B$ &    6.5      &   440        &     1170      &    0.64  \\
\hline\hline
\end{tabular}
\caption{The best fit of parameters in case A and case B.}
\label{parameter}
\end{center}
\end{table}

\begin{figure}
\begin{center}
\includegraphics[width=64mm,clip=true,keepaspectratio=true]{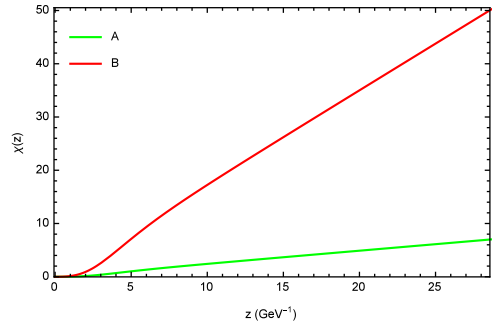}
\caption{The bulk scalar VEV $\chi(z)$ in case A and case B.}
\label{scalar VEV}
\end{center}
\end{figure}

The bulk scalar VEV $\chi(z)$, which can be solved from Eq.(\ref{eom-chi}) with the required UV and IR conditions, is shown in Fig.\ref{scalar VEV}, and the value of $\sigma$ can be extracted from the UV limit of $\chi(z)$. The chiral condensate is $\langle\bar{q}q\rangle=\sigma=(175~\mathrm{MeV})^3$ in case A and $\langle\bar{q}q\rangle=\frac{N_c}{4\pi^2}\sigma=(214~\mathrm{MeV})^3$ in case B. We remark here that the different formulas of chiral condensate in case A and case B are due to the different ways of realizing the correct $N_c$ scaling behavior of chiral condensate, for which one can refer to \cite{Colangelo:2011sr,Cherman:2008eh}.

\subsection{Meson spectrum}\label{meson spectrum}
\subsubsection{Pseudoscalar meson}
\begin{figure}
\begin{center}
\includegraphics[width=61.5mm,clip]{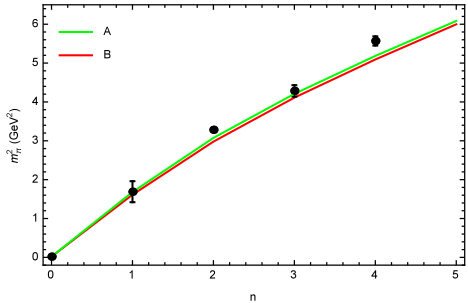}
\end{center}
\caption{The squared masses of pseudoscalar meson $m_{\pi}^2$ versus the radial excitation number $n$ in case A and case B. The black points with error bars are experimental data taken from \cite{Agashe:2014kda}.}
\label{pi-mass square}
\end{figure}

Now we are ready to calculate the meson spectrum, and the pseudoscalar meson will be firstly considered. To derive the EOM of pseudoscalar meson, we need to decompose the bulk gauge field $A_{\mu}^a$ into $A_{\mu}^a=A_{\mu \bot}^a+\partial_{\mu}\phi^a$, which eliminates the cross term of pseudo-scalar and axial-vector meson fields, but introduces a mixing between $\pi^a$ and $\phi^a$. Substituting $X=(\frac{\chi}{2}+S)e^{2i\pi}$ in the action (\ref{meson action}) and using the Kaluza-Klein (KK) decomposition $\pi^a(x,z)=\sum_n\varphi_n(x)\pi^a_n(z)$, the EOM of pseudoscalar meson can be derived, in the axial gauge $A_z=0$, as follows
\begin{align}
&\partial_z\left(e^{A-\Phi}\partial_z\phi_n^{a}\right)+g_5^2\,\chi^2 e^{3A-\Phi}(\pi_n^{a}-\phi_n^{a})=0, \label{pi meson1}\\
&m_n^2\,\partial_z\phi_n^{a}-g_5^2\,e^{2A}\chi^2\partial_z\pi_n^{a}=0. \label{pi meson2}
\end{align}
Eqs.(\ref{pi meson1}) and (\ref{pi meson2}) can be solved by the shooting method with the boundary conditions $\phi_n^a(z\rightarrow0)=0$,
$\partial_z \phi_n^a(z\rightarrow\infty)=0$ and $\pi_n^a(z\rightarrow0)=0$, then we obtain the mass spectrum of pseudoscalar mesons, which can be compared with the experimental results. The numerical calculations are shown in Table \ref{pi spectrum} and Fig.\ref{pi-mass square}. It can be seen that both case A and case B can give a good description of pseudoscalar meson spectrum which is consistent with the experiments.
\begin{table*}
\begin{center}
    \begin{tabular}{ccccccc}
       \hline\hline
       $\pi$ & 0 & 1 & 2 & 3 & 4 & 5 \\
       \hline\hline
       Exp. (MeV) & 139.6 & $1300\pm100$ & $1812\pm12$ & $2070\pm35$ & $2360\pm25$ & --- \\
       \hline
       A (MeV) & 139.6 & 1296 & 1753 & 2051 & 2277 & 2467  \\
       \hline
       B (MeV) & 138.6 & 1270 & 1726 & 2028 & 2257 & 2449  \\
       \hline\hline
     \end{tabular}
\caption{The theoretical results of the radial excited pseudoscalar meson spectrum in case A and case B. The experimental data are taken from \cite{Agashe:2014kda}.}
\label{pi spectrum}
\end{center}
\end{table*}

\subsubsection{Scalar meson}
Assuming $S(x,z)=\sum_n\mathcal{S}_n(x)S_n(z)$ and substituting it in the action (\ref{meson action}), the EOM of scalar meson can be derived as
\begin{equation}\label{f0 EOM1}
\small
\begin{split}
&\partial_z\left(e^{\omega_s(z)}\partial_z S_n(z)\right)-e^{2A(z)+\omega_s(z)}\left(m_5^2(z)+\frac{3}{2}\lambda \chi^2(z)\right)S_n(z)\\
&+e^{\omega_s(z)}m_n^2S_n(z)=0
\end{split}
\end{equation}
with $\omega_s(z)=3A(z)-\Phi(z)$. To simplify the above equation, let us define $S_n(z)=e^{-\omega_s(z)/2} s_n(z)$, then Eq.(\ref{f0 EOM1}) can be rewritten as the schr\"{o}dinger form:
\begin{equation}\label{f0 EOM2}
-s_n''(z)+V_s(z) s_n(z)-m_n^2 s_n(z)=0
\end{equation}
with the potential
\begin{equation}\label{f0 potential}
\begin{split}
V_s(z)=&\frac{1}{2}\omega_s''(z)+\frac{1}{4}\omega_s'(z)^2\\
&+e^{2A(z)}\left(m_5^2(z)+\frac{3}{2}\lambda \chi^2(z)\right).
\end{split}
\end{equation}

\begin{figure}
\begin{center}
\includegraphics[width=61.5mm,clip]{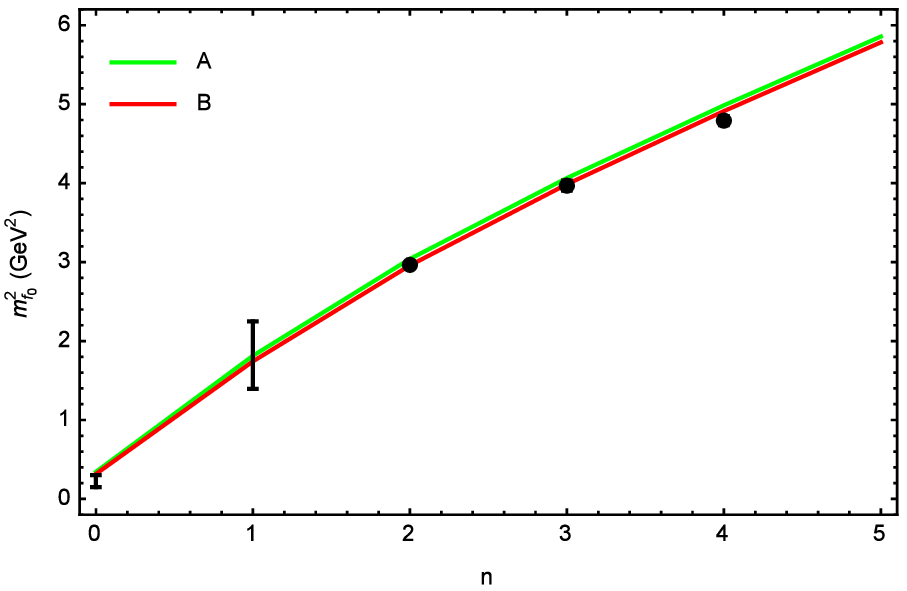}
\includegraphics[width=64mm,clip]{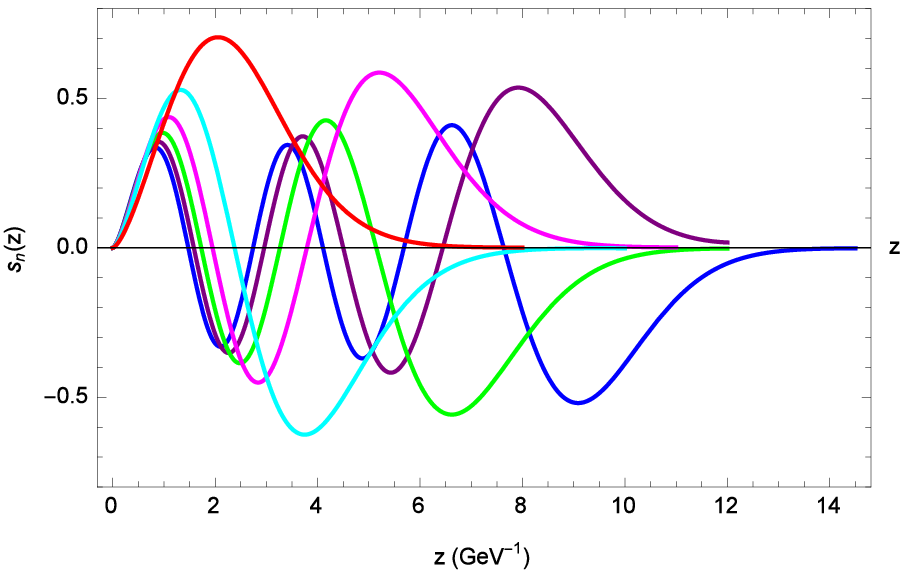}
\end{center}
\caption{The squared masses of scalar mesons $m_{f_0}^2$ versus the radial excitation number $n$ in case A and case B (top) and the corresponding wave functions $s_n(z)$ in case A (down). In the top panel, the black points with error bars are experimental data taken from \cite{Agashe:2014kda}.}
\label{f0-mass-wave}
\end{figure}

Using the shooting method with the boundary conditions $s_n(z\rightarrow0)=0$ and $s_n(z\rightarrow\infty)=0$, the mass spectrum of resonance scalar mesons can be calculated. We show the results in Table \ref{f0 spectrum} and Fig.\ref{f0-mass-wave}, where the scalar wave functions of the first six states in case A are also plotted.
\begin{table*}
    \begin{tabular}{ccccccc}
       \hline\hline
       $f_0$ & 0 & 1 & 2 & 3 & 4 & 5 \\
       \hline\hline
       Exp. (MeV) & $400-550$ & $1200-1500$ & $1722^{+6}_{-5}$ & $1992 \pm 16$ & $2189 \pm 13$ & --- \\
       \hline
         A (MeV) & 586 & 1346 & 1743 & 2016 & 2232 & 2420  \\
       \hline
         B (MeV) & 562 & 1320 & 1720 & 1997 & 2216 & 2405  \\
       \hline\hline
     \end{tabular}
\caption{The calculated results of the radial excited scalar meson spectrum in case A and case B. The experimental data are taken from \cite{Agashe:2014kda}.}
\label{f0 spectrum}
\end{table*}

We remark that in our calculation the mass spectrum of radial excited scalar states is identified as the one of $SU(3)$ singlet scalar mesons. As there are many uncertainties for the scalar meson states in the experiment \cite{Agashe:2014kda}, our selection of relevant scalar mesons is based on the previous studies which involve theoretical inferences \cite{Sui:2009xe}. Once we choose the proper scalar mesons, the model calculation and experimental data show a good consistency, as can be seen in Fig.\ref{f0-mass-wave}.

\subsubsection{Vector meson}
\begin{figure}
\begin{center}
\includegraphics[width=61.5mm,clip]{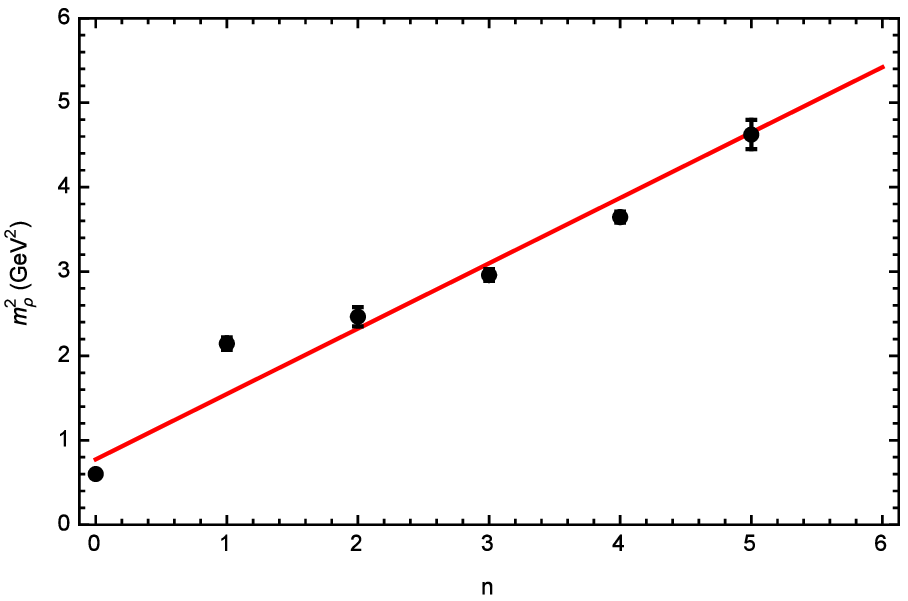}
\includegraphics[width=64mm,clip]{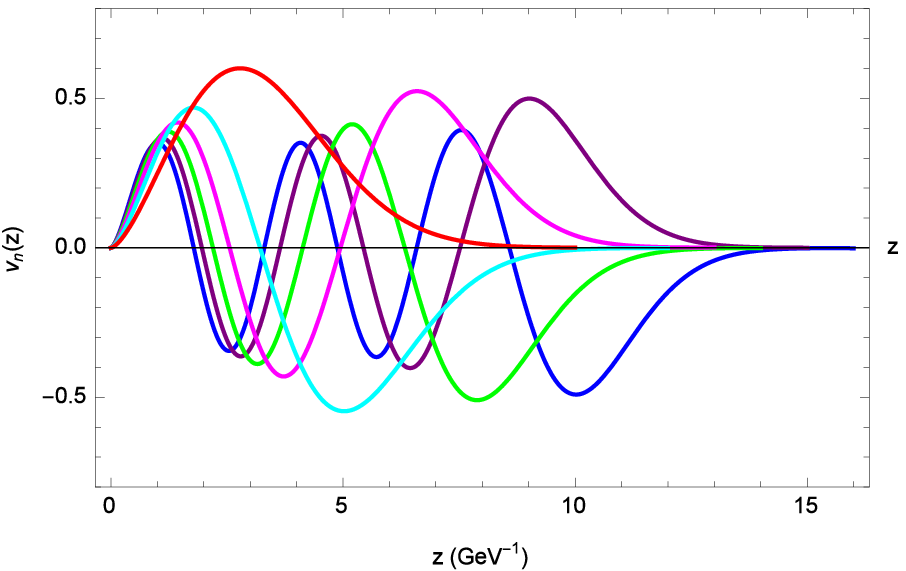}
\end{center}
\caption{The squared vector meson masses $m_{\rho}^2$ versus the radial excitation number $n$ (top) and the corresponding wave functions $v_n(z)$ in case A (down). In the top panel, the black points with error bars are experimental data \cite{Agashe:2014kda} and the red line denotes the theoretical calculations in both case A and case B.}
\label{v-mass-wave}
\end{figure}

Following the same procedures as before, let us give the EOM of vector meson which can be derived from the action (\ref{meson action}) in the axial gauge $V_5=0$:
\begin{equation}\label{vector EOM1}
\partial_z(e^{A(z)-\Phi(z)}\partial_zV_n(z))+m_n^2e^{A(z)-\Phi(z)}V_n(z)=0.
\end{equation}
Defining $V_n(z)=e^{\omega_{v}(z)/2} v_n(z)$ with $\omega_v(z)=\Phi(z)-A(z)$, the above equation can be transformed into the schr\"{o}dinger form:
\begin{equation}\label{vector EOM2}
\begin{split}
&-v_n''(z)+\left(\frac{1}{4}\omega_{v}'(z)^2-\frac{1}{2}\omega_{v}''(z)\right) v_n(z)\\
&-m_n^2 v_n(z)=0,
\end{split}
\end{equation}
from which we see that the vector meson EOM only includes the parameter $\mu_g$, which is the same as the original soft-wall model \cite{Karch:2006pv}, and analytical solutions can be easily obtained in this case. Numerically, Eq.(\ref{vector EOM2}) can be solved by the shooting method with the boundary conditions $v_n(z\rightarrow0)=0$ and $v_n(z\rightarrow\infty)=0$. The numerical results of meson spectrum and corresponding wave functions are shown in Table \ref{vector spectrum} and Fig.\ref{v-mass-wave}, where we can see that the theoretical results are in accord with experimental data. As the same value of $\mu_g$ is used in case A and case B, there is no difference for the vector meson spectrum in the two cases.
\begin{table*}
    \begin{tabular}{cccccccc}
       \hline\hline
       $\rho$ & 0 & 1 & 2 & 3 & 4 & 5 & 6 \\
       \hline\hline
       Exp. (MeV) & $775.26\pm0.25$ & $1465\pm25$ & $1570\pm 36$ & $1720\pm 20$ & $1909\pm17$ & $2150\pm40$ & --- \\
       \hline
       A/B (MeV) & 880 & 1245 & 1524 & 1760 & 1968 & 2156 & 2328 \\
       \hline\hline
     \end{tabular}
\caption{The experimental data and model calculation of radial excited vector meson spectrum. The experimental data are taken from \cite{Agashe:2014kda}.}
\label{vector spectrum}
\end{table*}

\subsubsection{Axial-vector meson}
In the $A_z=0$ gauge, the EOM of axial-vector meson, which is represented by the transverse axial gauge field $A_{\mu \bot}^a$, can be derived from the action (\ref{meson action}) as
\begin{equation}\label{av EOM1}
\begin{split}
& e^{\Phi(z)}\partial_z(e^{A(z)-\Phi(z)}\partial_z A_{n}(z))+ m_n^2 e^{A(z)} A_{n}(z)\\
 &-g_5^2 e^{3A(z)} \chi^2(z) A_{n}(z)=0.
\end{split}
\end{equation}
Again, by defining $A_n(z)=e^{\omega_{v}(z)/2} a_n(z)$, the above equation can be rewritten as
\begin{equation}\label{av EOM2}
\begin{split}
&-a_n''(z)+\left(\frac{1}{4}\omega_{v}'(z)^2-\frac{1}{2}\omega_{v}''(z)+g_5^2\,e^{2A(z)}\chi^2(z)\right) a_n(z)\\
&-m_n^2 a_n(z)=0.
\end{split}
\end{equation}
Compared with the EOM of vector meson Eq.(\ref{vector EOM2}), there is an additional term $g_5^2\,e^{2A(z)}\chi^2(z)$ in the above equation, which contributes to the mass split of chiral partners due to the chiral symmetry breaking pattern in the highly excited mesons which is supposed in QCD \cite{Shifman:2007xn}. In this case, we need to require this additional term to approach a nonzero constant in the IR limit, which in turn entails linear IR asymptotics of the bulk scalar VEV $\chi(z)$, as has been noted above.

\begin{figure}
\begin{center}
\includegraphics[width=61.5mm,clip]{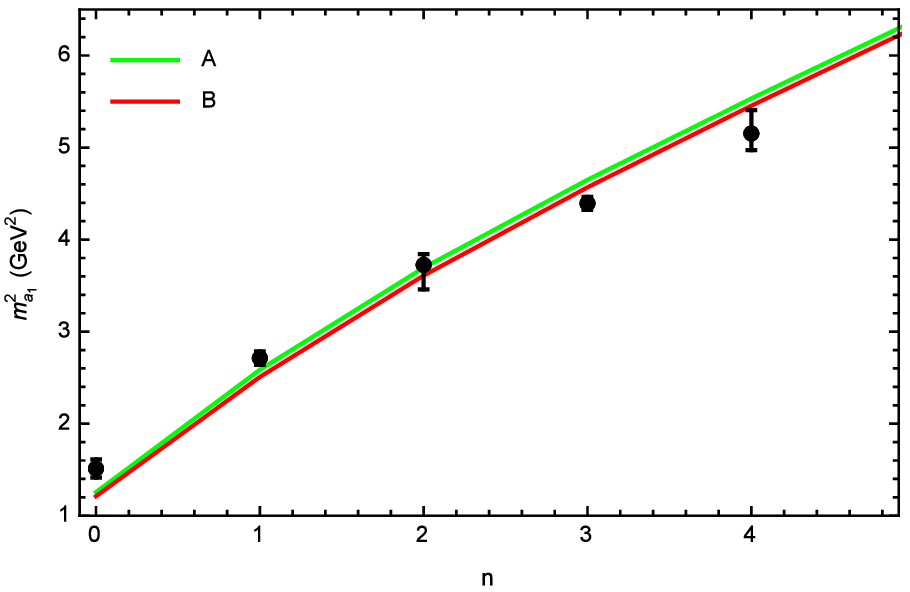}
\includegraphics[width=64mm,clip]{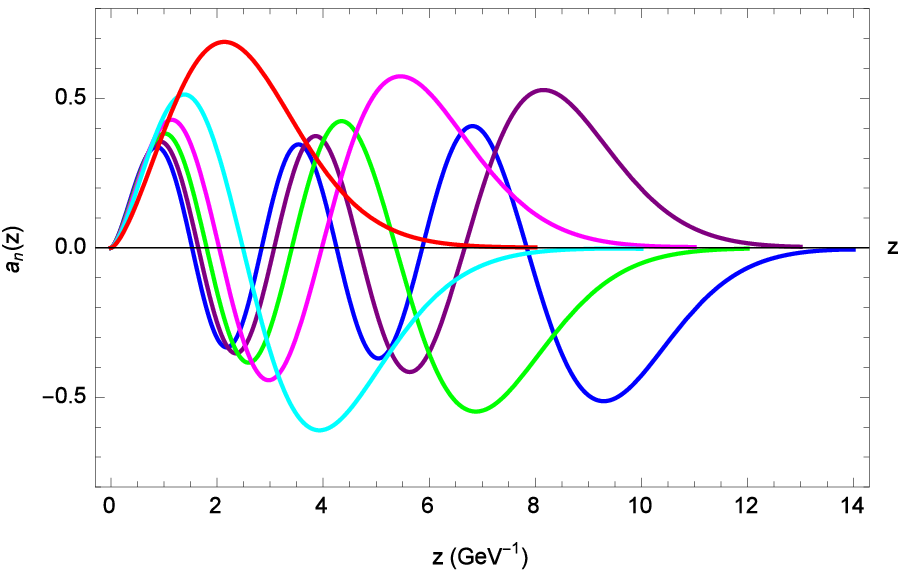}
\end{center}
\caption{The squared masses of axial-vector meson $m_{a_1}^2$ versus the radial excitation number $n$ in case A and case B (top) and the corresponding wave functions $a_n(z)$ in case A (down). In the top panel, the black points with error bars are experimental data taken from \cite{Agashe:2014kda}.}
\label{av-mass-wave}
\end{figure}

Similarly, the Eq.(\ref{av EOM2}) can be solved with the boundary conditions $a_n(z\rightarrow0)=0$ and $a_n(z\rightarrow\infty)=0$. The numerical results are presented in Table \ref{av spectrum} and Fig.\ref{av-mass-wave}, from which one can see that the results in case A and case B agree well with the experimental data.
\begin{table*}
     \begin{tabular}{ccccccc}
       \hline\hline
       $a_1$ & 0 & 1 & 2 & 3 & 4 & 5 \\
       \hline\hline
       Exp. (MeV) & $1230\pm40$ & $1647\pm22$ & $1930^{+30}_{-70}$ & $2096\pm17$ & $2270^{+55}_{-40}$ & --- \\
       \hline
         A (MeV) & 1121 & 1608 & 1922 & 2156 & 2352 & 2526  \\
      \hline
         B (MeV) & 1101 & 1584 & 1900 & 2137 & 2335 & 2511  \\
      \hline\hline
     \end{tabular}
\caption{The theoretical results of radial excited axial-vector meson spectrum in case A and case B. The experimental data are taken from \cite{Agashe:2014kda}.}
\label{av spectrum}
\end{table*}

\subsection{Pion form factor, $\pi-\rho$ coupling constant and decay constants}\label{couplings}
From the above calculations, one can see that our improved soft-wall AdS/QCD model can give a good description for the light meson spectra, and the results calculated in case A and case B have little differences. For further test of the model in case A and case B, we compute the pion form factor $F_{\pi}(Q^2)$, the $\pi-\rho$ coupling constant $g_{\rho\pi\pi}$ and the decay constants of $\pi$, $\rho$ and $a_1$ mesons.

\begin{figure}
\begin{center}
\includegraphics[width=64mm,clip]{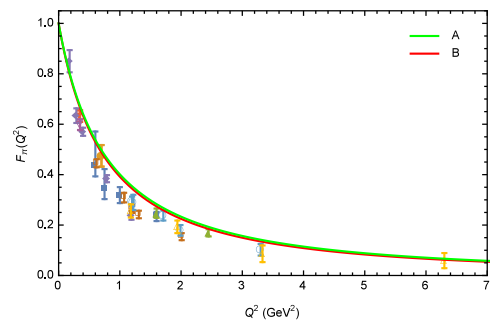}
\caption{The solid line shows the space-like pion form factor $F_\pi(q^2)$ predicted in case A and case B. The points are the data of experiments. The square: the Jefferson Lab $F_{\pi }$ Collaboration \cite{Tadevosyan:2007yd[1]}. The circle: data from P. Brauel, et al. \cite{Brauel:1979zk[10]}, reanalyzed by the Jefferson Lab $F_{\pi }$ Collaboration \cite{Tadevosyan:2007yd[1]}. The up triangle: the Jefferson Lab $F_{\pi }$-2 Collaboration \cite{Horn:2006tm[2]},the Jefferson Lab $F_{\pi }$ Collaboration \cite{Huber:2008id[3]}. The down triangle: data from H. Ackermann, et al. \cite{Ackermann:1977rp[8]}, reanalyzed by The Jefferson Lab $F_{\pi }$ Collaboration \cite{Huber:2008id[3]}. The filled diamond: CEA data \cite{Brown:1973wr[4]}, analysed by C. J. Bebek et al. \cite{Bebek:1977pe[7]}. The empty square: Earlier Cornell data \cite{Bebek:1974iz[5]}, analysed by C. J. Bebek et al. \cite{Bebek:1977pe[7]}. The empty circle: Later Cornell data \cite{Bebek:1974ww[6]}, analysed by C. J. Bebek et al. \cite{Bebek:1977pe[7]}. The empty up triangle: C.J. Bebek, et al. \cite{Bebek:1977pe[7]}. The empty down triangle: H. Ackermann, et al. \cite{Ackermann:1977rp[8]}.}
\label{fig form factor}
\end{center}
\end{figure}

The pion form factor $F_{\pi}(Q^2)$ can be extracted in our setup from the cubic terms of the 5D action \cite{Grigoryan:2007wn,Kwee:2007nq}:
\begin{equation}\label{pion form factor}
\small
F_{\pi}(Q^2)= \frac{1}{N}\int dz\, e^{A-\Phi}V(q,z)\left(\frac{(\partial_z\phi)^2}{g_5^2}+e^{2A}\chi^2(\pi-\phi)^2\right),
\end{equation}
where $Q^2\equiv-q^2>0$, $\pi(z)$ and $\phi(z)$  are the ground state wave functions of pseudoscalar meson, $V(q,z)$ is the vector bulk-to-boundary propagator satisfying
\begin{equation}\label{propagator}
\partial_z(e^{A-\Phi}\partial_z V(q,z))+q^2 e^{A-\Phi} V(q,z)=0
\end{equation}
with the boundary conditions $V(q,0)=1$ and $\partial_z V(q,\infty)=0$, and the normalization constant $N$ of the form
\begin{equation}\label{Normalize form factor}
N = \int dz\, e^{A-\Phi}\left(\frac{(\partial_z\phi)^2}{g_5^2}+e^{2A}\chi^2(\pi-\phi)^2\right),
\end{equation}
which guarantees $F_{\pi}(0)=1$. The numerical calculations for the pion form factor in case A and case B are presented in Fig.\ref{fig form factor}, from which one can see that the numerical results have a good agreement with the experimental data and there is no obvious deviation of the results in case A and case B.

The $\pi-\rho$ coupling constant $g_{\rho\pi\pi}$ can also be derived from the cubic terms of action (\ref{meson action}) as \cite{Erlich:2005qh,Grigoryan:2007wn,Kwee:2007nq}
\begin{equation}\label{vector coupling}
g_{\rho\pi\pi} = \frac{g_5}{N}\int dz\,e^{A-\Phi}V_{\rho}(z)\left(\frac{(\partial_z\phi)^2}{g_5^2}+e^{2A}\chi^2(\pi-\phi)^2\right),
\end{equation}
where $V_{\rho}(z)$ is the ground state wave function of vector meson satisfying the normalization condition $\frac{1}{k}\int dz\,e^{A-\Phi}V_{\rho}^2(z)=1$ with $k=1$ for case A and $k=\frac{16\pi^2}{3}$ for case B, and $N$ the same as that in Eq.(\ref{pion form factor}).

The decay constants $f_{\pi}$, $F_{\rho}$ and $F_{a_1}$ can be extracted from the two-point correlation functions of relevant currents as follows \cite{Erlich:2005qh}
\begin{align}\label{decay constant}
f_{\pi}^2  &=-\frac{1}{k g_5^2}e^{A-\Phi}\partial_z A(0,z)|_{z\to0},\\
F_{\rho}^2 &=\frac{1}{k g_5^2}\left(e^{A-\Phi}\partial_z V_{\rho}(z)|_{z\to0}\right)^2,\\
F_{a_1}^2  &=\frac{1}{k g_5^2}\left(e^{A-\Phi}\partial_z A_{a_1}(z)|_{z\to0}\right)^2,
\end{align}
where $V_{\rho}(z)$ and $A_{a_1}(z)$ are the ground state wave functions of vector and axial-vector mesons normalized by $\frac{1}{k}\int dz\,e^{A-\Phi}V_{\rho}^2(z)=1$ and $\frac{1}{k}\int dz\,e^{A-\Phi}A_{a_1}^2(z)=1$, and $A(0,z)$ is the axial-vector bulk-to-boundary propagator at zero momentum, which satisfies the differential equation $\partial _z\left(e^{A_s-\Phi }\left(\partial _zA(0,z)\right)\right)-g_5^2e^{3 A_s-\Phi }\chi ^2A(0,z)=0$ with boundary conditions $A(0,0)=1$ and $\partial_z A(0,\infty)=0$.

We present the numerical calculations for the $\pi-\rho$ coupling constant $g_{\rho\pi\pi}$ and the decay constants $f_{\pi}$, $F_{\rho}$ and $F_{a_1}$ in Table \ref{coupling-decay}, from which we see that the numerical results of $g_{\rho\pi\pi}$ and $f_{\pi}$ in case A and case B are close to each other, while there are notable differences between these two cases for the results of $F_{\rho}$ and $F_{a_1}$, which result from the different normalization conditions of the $\rho~(a_1)$ wave functions due to the different values of $k$ used in the two cases. Obviously, the results of $F_{\rho}$ and $F_{a_1}$ calculated in case A are closer to experiments than the ones in case B, though there are still $15\%$ deviation from the experimental data for $F_{\rho}^{1/2}$ and $10\%$ deviation for $F_{a_1}^{1/2}$ in case A. The model predictions for $g_{\rho\pi\pi}$ are little larger than the experimental results, while the values of $f_{\pi}$ are too small when compared with experiment. It should be noted that the Gell-Mann-Oakes-Renner (GOR) relation $f_\pi^2m_\pi^2=2m_q\sigma$ still holds for the reason that this relation is an intrinsic one which can be derived from the soft-wall AdS/QCD model per se. The small value of $f_{\pi}$ in our setup is due to the small value of $\sigma$ extracted from the fitting of light meson spectra.
\begin{table}
\begin{tabular}{ccccc}
\hline\hline
     &  $g_{\rho\pi\pi}$(MeV) & $f_{\pi}$(MeV) & $F_{\rho}^{1/2}$(MeV)  & $F_{a_1}^{1/2}$(MeV) \\
\hline
 Exp. &     $6.03\pm0.07$     &  $92.4\pm0.35$  &  $346.2\pm1.4$  &  $433\pm13$  \\
 \hline
   A &    8.38     &   43.3        &     296      &    389  \\
\hline
   B &    8.79     &   40.9       &     796      &    1025  \\
\hline\hline
\end{tabular}
\caption{The numerical results of $g_{\rho\pi\pi}$, $f_{\pi}$, $F_{\rho}$ and $F_{a_1}$ in case A and case B. The experimental data are taken from \cite{Erlich:2005qh}.}
\label{coupling-decay}
\end{table}

\subsection{Relaxing the scalar spectrum}\label{relax-scalar}
In the above study, we compute the light meson spectra which have a good agreement with data of experiments. However, the other quantities such as the $\pi-\rho$ coupling constant and the decay constants do not fit the experimental values well, especially the pion decay constant $f_{\pi}$, which is too small in both case A and case B. On the other hand, it seems that the case A with $k=1$ gives better results than the case B with $k=\frac{16\pi^2}{N_c}$, in consideration of the (axial-)vector decay constant calculated above.

As the scalar mesons have so many uncertainties in the experiment, we now relax the scalar part and only fit the spectra of vector, axial-vector and pseudoscalar mesons in case A with another consideration of the chiral phase transition which will be addressed later. The way of mass-spectrum fitting is the same as before, with the best-fitting parameters listed in Table \ref{parameter2}. Note that $k=1$ and $\zeta=\frac{\sqrt{N_c}}{2\pi}$, as mentioned in Sec.\ref{model}.
\begin{table}
\begin{center}
\begin{tabular}{ccccc}
\hline\hline
Case&  $m_q$(MeV) & $\mu_g$(MeV) & $\mu_c$(MeV)  & $\lambda$ \\
\hline
   A &    3.22     &   440        &     1450      &    80  \\
\hline\hline
\end{tabular}
\caption{The best fit of parameters in case A without consideration of scalar meson spectrum.}
\label{parameter2}
\end{center}
\end{table}

From the profile of bulk scalar VEV $\chi(z)$, the chiral condensate can be extracted to be $\langle\bar{q}q\rangle=\sigma=(247~\mathrm{MeV})^3$, which is much larger than the one calculated in Sec.\ref{parameter-fit}. As the value of $\mu_g$ does not change, the vector meson spectrum is the same as that in Table \ref{vector spectrum}. The numerical calculations for the mass spectra of pseudoscalar and axial-vector mesons are presented in Table \ref{pi-av spectrum}, which shows good consistency with experiments.
\begin{table*}
    \begin{tabular}{ccccccc}
       \hline\hline
                       & 0 & 1 & 2 & 3 & 4 & 5 \\
       \hline\hline
       $\pi$ Exp.(MeV) & 139.6 & $1300\pm100$ & $1812\pm12$ & $2070\pm35$ & $2360\pm25$ & --- \\
       \hline
       A (MeV)  & 139.7 & 1510 & 1835 & 2061 & 2254 & 2429  \\
       \hline
       $a_1$ Exp.(MeV) & $1230\pm40$ & $1647\pm22$ & $1930^{+30}_{-70}$ & $2096\pm17$ & $2270^{+55}_{-40}$ & --- \\
       \hline
       A (MeV)  & 1310 & 1686 & 1924 & 2126 & 2306 & 2472  \\
       \hline\hline
     \end{tabular}
\caption{The calculated mass spectra of pseudoscalar and axial-vector mesons in case A without consideration of scalar meson spectrum. The experimental data are taken from \cite{Agashe:2014kda}.}
\label{pi-av spectrum}
\end{table*}

Then we calculate the $\pi-\rho$ coupling constant $g_{\rho\pi\pi}$ and the decay constants of $\pi$, $\rho$ and $a_1$ mesons again with the fitting parameters in Table \ref{parameter2}, and the numerical results are shown in Table \ref{coupling-decay2}, from which we can see that the numerical calculation for the $\pi-\rho$ coupling constant and decay constants in this case fits experimental data much better than that shown in Table \ref{coupling-decay}. The value of $f_{\pi}$ has increased from about $40~\mathrm{MeV}$ to about $70~\mathrm{MeV}$. The value of $g_{\rho\pi\pi}$ is also closer to the experimental result than that in Table \ref{coupling-decay}, and the calculated axial-vector decay constant $F_{a_1}$ is almost the same as the experimental data. Furthermore, as we will see later, the behavior of chiral thermal transition and the chiral transition temperature in this case are very consistent with the lattice simulations.
\begin{table}
\begin{tabular}{ccccc}
\hline\hline
     &  $g_{\rho\pi\pi}$(MeV) & $f_{\pi}$(MeV) & $F_{\rho}^{1/2}$(MeV)  & $F_{a_1}^{1/2}$(MeV) \\
\hline
 Exp. &     $6.03\pm0.07$     &  $92.4\pm0.35$  &  $346.2\pm1.4$  &  $433\pm13$  \\
\hline
 A &    4.88     &   70.7        &     296      &    432  \\
\hline\hline
\end{tabular}
\caption{The values of $g_{\rho\pi\pi}$, $f_{\pi}$, $F_{\rho}$ and $F_{a_1}$ in case A without consideration of scalar meson spectrum. The experimental data are taken from \cite{Erlich:2005qh}.}
\label{coupling-decay2}
\end{table}

\section{Chiral thermal transition}\label{chiraltransition}
Now we turn to the finite temperature case and investigate the thermal behavior of chiral condensate, which has attracted much attention in the study of QCD phase transitions. In the two-flavor case, lattice simulations \cite{Laermann:2003cv,Kanaya:2010qd} indicate that chiral thermal transition at physical quark mass is a crossover one which happens at around $T_c\simeq170~\mathrm{MeV}$. In the holographic framework, there have been many studies on chiral phase transition both in the top-down approaches and in the bottom-up approaches \cite{Colangelo:2011sr,Erdmenger:2007cm,Kim:2012ey,Jarvinen:2011qe,Alho:2012mh,Chelabi:2015cwn,Chelabi:2015gpc}. Few of them can realize the crossover behavior of chiral transition, except \cite{Chelabi:2015cwn,Chelabi:2015gpc}, where it has been shown that a quartic term of bulk scalar together with a modified dilation play a key role in producing the correct behavior of chiral phase transition. In this work, the simplest $z^2$ dilaton profile has been used, but with a modified bulk scalar mass which can be well motivated from the quark mass anomalous dimension. It will be shown that the correct chiral transition behavior can also be realized in our setup.

To introduce temperature in our setup, we simply use an AdS-Schwarzchild black hole background with the metric ansatz:
\begin{equation}\label{BH metric}
ds^2=e^{2A(z)}\left(f(z)dt^2-dx_idx^i-\frac{dz^2}{f(z)}\right)
\end{equation}
with
\begin{equation}
f(z)=1-\frac{z^4}{z_h^4},
\end{equation}
where $z_h$ is the horizon of black hole which is related to the Hawking temperature $T$ by the formula
\begin{equation}\label{hawking T}
T=\frac{1}{4\pi}\left|\frac{df}{dz}\right|_{z_{h}}=\frac{1}{\pi z_{h}}.
\end{equation}

\begin{figure}[H]
\begin{center}
\includegraphics[width=64mm,clip=true,keepaspectratio=true]{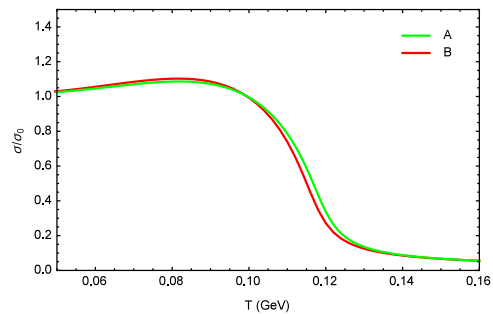}
\caption{Temperature dependent behavior of the rescaled chiral condensate $\frac{\sigma}{\sigma_0}$ in case A and case B. Note that $\sigma_0=\sigma(0)$.}
\label{chiral-T}
\end{center}
\end{figure}

As was usually done, the Hawking temperature $T$ is identified as the temperature of QCD. To extract the finite temperature behavior of chiral condensate, we need to solve the EOM of the bulk scalar VEV $\chi(z)$, which can be derived from the action (\ref{meson action}) with the black hole metric as
\begin{equation}\label{eom-chi-T}
\begin{split}
&\chi''(z)+\left(3A'(z)-\Phi'(z)+\frac{f'(z)}{f(z)}\right)\chi'(z)\\
&-\frac{e^{2A(z)}}{f(z)}\left(m_5^2\,\chi(z)+\frac{\lambda}{2}\chi^3(z)\right)=0.
\end{split}
\end{equation}
In the finite temperature case, the UV asymptotic solution of the above equation still has the form
\begin{equation}\label{chi-UV-T}
\chi(z\sim 0)=m_q\,\zeta\,z+\frac{\sigma(T)}{\zeta}\,z^3+\cdots
\end{equation}
with a temperature dependent chiral condensate $\sigma(T)$, which can be obtained by solving Eq.(\ref{eom-chi-T}) with
a regular condition imposed at the horizon of the black hole \cite{Colangelo:2011sr,Chelabi:2015gpc}. Using the same values of parameters listed in Table \ref{parameter}, the numerical calculations for $\sigma(T)$ in case A and case B are shown in Fig.\ref{chiral-T}, from which we see that the crossover behavior of chiral thermal transition does manifest in our improved soft-wall AdS/QCD model. However, the transition temperature $T_c$ is within $110\sim120~\mathrm{MeV}$, much smaller than the expected value $T_c\simeq170~\mathrm{MeV}$ from the lattice simulations \cite{Laermann:2003cv}. Besides that, there is a temperature region below $T_c$ where $\sigma$ rises up gradually by a small value as $T$ increases, which is unreasonable physically. The behavior is the same as that in \cite{Colangelo:2011sr}, where a bump appears near the chiral transition region.

\begin{figure}
\begin{center}
\includegraphics[width=64mm,clip=true,keepaspectratio=true]{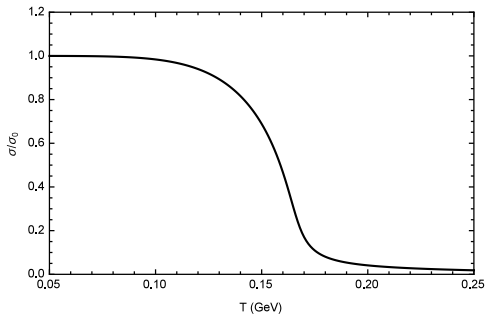}
\caption{Temperature dependent behavior of the rescaled chiral condensate $\frac{\sigma}{\sigma_0}$ in case A without consideration of scalar meson spectrum. Note that $\sigma_0=\sigma(0)$.}
\label{chiral-T2}
\end{center}
\end{figure}

As has been seen in Sec.\ref{relax-scalar}, we get more consistent results both for the meson spectra and for the other low-energy characteristic quantities such as the $\pi-\rho$ coupling constant and the decay constants when the scalar meson spectrum is not considered. For comparison, let us also give the result of chiral thermal transition in case A without consideration of scalar meson spectrum, which is shown in Fig.\ref{chiral-T2}. We see that the strange rising-up behavior of chiral condensate has disappeared and a transition temperature around $T_c\simeq170~\mathrm{MeV}$ is attained. It is remarkable that a perfectly consistent chiral transition behavior is also obtained in this case as we only relax the scalar part which is messy in the experiments.

We remark that a rescaled chiral condensate $\frac{\sigma}{\sigma_0}$ with $\sigma_0=\sigma(0)$ has been used in our comparison of the results of case A and case B for the different definitions of chiral condensate in these two cases. Indeed, the chiral condensate in the holographic framework admits another indeterminacy, which can be seen easily from the UV solution of $\chi(z)$:
\begin{equation}\label{chi-UV2}
\small
\begin{split}
&\chi(z\sim 0) \\
=&m_q z+\sigma(T) z^3+\left(m_q\mu_g^2+\frac{m_q^3\lambda}{4}-\frac{m_q\mu_c^2}{2}\right)z^3\mathrm{log}(\alpha z)+\cdots\\
=&m_q z+\left[\sigma(T)+\left(m_q\mu_g^2+\frac{1}{4}m_q^3\lambda-\frac{1}{2}m_q\mu_c^2\right)\mathrm{log}\alpha\right]z^3+\cdots
\end{split}
\end{equation}
with $\alpha$ an arbitrary constant (we have set $\zeta=1$ for convenience). To tackle this problem, a full consideration might involve the theory of holographic renormalization. However, as the $\mathrm{log}(\alpha z)$ term does not depend on temperature, this uncertainty would not affect our investigation of the chiral thermal transition.

\section{Summary and conclusions}\label{conclusion}


In this paper, we consider a simply improved soft-wall AdS/QCD model with a quartic term of bulk scalar and a modified 5D conformal mass of the bulk scalar field based on our previous work \cite{Cui:2016ocl}. The modified bulk scalar mass $m_5(z)$ is well motivated from the quark mass anomalous dimension, and its form is constrained by the UV and IR asymptotics of the bulk scalar VEV $\chi(z)$. In consideration of all this, the simplest parameterization of $m_5(z)$ has been used in our model. It is found that the spontaneous chiral symmetry breaking can be realized reasonably in this simply improved soft-wall model and the correct behavior of chiral thermal transition is also obtained.

The two cases (case A and case B) with different values of $k$ in the action are analyzed: the normal way is taking $k=1$, as did in the previous works including the original hard-wall and soft-wall models \cite{Erlich:2005qh,Karch:2006pv,Gherghetta:2009ac,Sui:2009xe,Li:2012ay,Cui:2016ocl}; another way was taken in \cite{Colangelo:2008us,Krikun:2008tf} where $k$ was fixed by correlators of scalar or pseudoscalar currents. It should be noted that the values of $k$ fixed in \cite{Colangelo:2008us,Krikun:2008tf} are different from each other, so some inconsistencies indeed exist in this way of fixing $k$. What is more, different values of $k$ will directly affect the calculations of the low energy quantities such as the $\pi-\rho$ coupling constant and decay constants, which might indicate further investigations on this issue.

The mass spectra of light mesons are calculated, which have a good agreement with experiments. Using the parameters fixed by the light meson spectra, we also calculate the pion form factor, the $\pi-\rho$ coupling constant and the decay constants of $\pi$, $\rho$ and $a_1$ mesons. It is found that the decay constants of $\rho$ and $a_1$ in case B are much larger than the experimental data, which can be attributed to the normalization condition of $\rho$ $(a_1)$ wave function in this case, while the decay constants of $\pi$ are too small in both case A and case B when compared with experiments. As the GOR relation still holds, the small $f_{\pi}$ results from the small value of $\sigma$ extracted from the UV limit of $\chi(z)$.

The thermal behavior of chiral condensate has also been studied on the basis of the zero-temperature case by introducing an AdS-Schwarzchild black hole background. It is found that the simply improved soft-wall model incorporates the crossover behavior of chiral thermal transition, which is not the case in the original soft-wall model \cite{Colangelo:2011sr}. However, the chiral transition temperature $T_c$ is calculated to be within $110\sim120~\mathrm{MeV}$, which is much smaller than the value $T_c\simeq170~\mathrm{MeV}$ indicated from lattice simulations \cite{Laermann:2003cv}. What is more, another unfavorable behavior still persist, as in the original soft-wall model, i.e., the slowly rising-up behavior of chiral condensate with increasing temperature in a region below $T_c$. We remark that this undesirable behavior is not inevitable in our model, but is related to the parameter $\mu_c$ of the bulk scalar mass. When $\mu_c$ is large enough, $\sigma$ will always go down with increasing temperature.

As an attempt, we have also made another fitting in case A without considering the scalar meson spectrum for the messy situation of scalar sector in experimental measurements. In this case, we find that the results of the $\pi-\rho$ coupling constant and the decay constants of $\pi$, $\rho$ and $a_1$ are much more consistent with experimental data. The chiral transition temperature is shown to be around $T_c\simeq170~\mathrm{MeV}$, and the bump existing in the former case disappears. However, as we relax the fitting of scalar spectrum, the mass of scalar meson in the ground state reaches to about $1000~\mathrm{MeV}$, which is much larger than the experimental value $400-550~\mathrm{MeV}$. The possible ways to reconcile the scalar meson part with other aspects might need further improvements of the model, such as the metric form \cite{Sui:2009xe}, the dilaton profile or the bulk scalar mass \cite{Cui:2016ocl} etc., which we do not address in this work. What is interesting for us is that we show possible ways to coordinate the meson spectrum and other low-energy characteristic quantities of QCD with the correct chiral phase structure in a simple soft-wall framework, which can be traced back to other theoretical ingredients (e.g., the quark mass anomalous dimension) of QCD.

Finally, we would like to remark that in our model two energy scales have been introduced by the parameters $\mu_g$ and $\mu_c$. The parameter $\mu_g=440~\mathrm{MeV}$ is close to $\Lambda_{\mathrm{QCD}}$ energy scale and determines the linear-confining property of meson spectrum, while the parameter $\mu_c\simeq 1.2~\mathrm{GeV}$ is closely related to the phenomena of chiral symmetry breaking. Indeed, there have been many discussions for the two different energy scales of QCD confinement and chiral symmetry breaking \cite{DW, Shuryak:1988ck}. Our holographic calculations in the soft-wall framework might be considered as some support for these facts.

\section{Acknowledgements}
The authors would like to thank Danning Li and Kaddour Chelabi for useful discussions. This work is supported in part by the National Nature Science Foundation of China (NSFC) under Grants No.10975170, No.10905084, No.10821504 and the Project of Knowledge Innovation Program (PKIP) of the Chinese Academy of Science.

\bibliographystyle{apsrev4-1} 
\bibliography{xampl} 


\end{document}